\documentclass[comment,prl,twocolumn,nofootinbib, preprintnumbers, superscriptaddress]{revtex4-2}
\usepackage{amsmath,amssymb,slashed,braket}
\usepackage{graphicx}
\usepackage{cancel}
\usepackage{epstopdf}
\usepackage{float}
\usepackage[colorlinks=true,
            linkcolor=red,
            urlcolor=red,
            citecolor=blue,
            bookmarks=true,
            bookmarksnumbered=true,
            breaklinks=true,
            pdfpagemode=Fullscreen,
            pdfstartview=FitBH]{hyperref}
\usepackage[normalem]{ulem}
\usepackage{subfig,ragged2e,tabularx}
\usepackage[normalem]{ulem}
\usepackage{comment}

\usepackage{color}
\definecolor{Orange}{cmyk}{0,0.61,0.87,0}
\definecolor{JungleGreen}{cmyk}{0.99,0,0.52,0}
\definecolor{OliveGreen}{cmyk}{0.64,0,0.95,0.40}
\definecolor{Brown}{cmyk}{0,0.81,1,0.60}
\definecolor{RoyalBlue}{cmyk}{0.71,0.53,0,0.12}
\definecolor{Gray}{cmyk}{0,0,0,0.40}
\definecolor{LightPink}{cmyk}{0.0,0.25,0,0}
\definecolor{LLightPink}{cmyk}{0.0,0.10,0,0}
\definecolor{LightBlue}{cmyk}{0.25,0,0,0}
\definecolor{LightGray}{cmyk}{0,0,0,0.2}

\usepackage{xcolor}
\definecolor{gesfpurple}{rgb}{0.47,0.19,0.42}

\definecolor{gesflanse}{rgb}{0.00,0.50,0.50}

\definecolor{gesfblue}{rgb}{0.08,0.42,0.76}

\definecolor{gesfred}{rgb}{1,0,0}

\definecolor{gesfwhite}{rgb}{1,1,1}

\definecolor{gesfblack}{rgb}{0,0,0}

\newcommand{\mc}{\mathcal}

\makeatletter
\renewcommand\normalsize{%
    \@setfontsize\normalsize{12pt}{14pt} 
}

\graphicspath{{Figures/}}

\begin{document}

\title{Quantum state tomography with muons}

\author{Leyun \surname{Gao}}
\email[]{seeson@pku.edu.cn}
\affiliation{State Key Laboratory of Nuclear Physics and Technology, School of Physics, Peking University, Beijing, 100871, China}

\author{Alim \surname{Ruzi}}
\email[]{alim.ruzi@pku.edu.cn}
\affiliation{State Key Laboratory of Nuclear Physics and Technology, School of Physics, Peking University, Beijing, 100871, China}

\author{Qite \surname{Li}}
\email[]{liqt@pku.edu.cn}
\affiliation{State Key Laboratory of Nuclear Physics and Technology, School of Physics, Peking University, Beijing, 100871, China}

\author{Chen \surname{Zhou}}
\email[]{czhouphy@pku.edu.cn}
\affiliation{State Key Laboratory of Nuclear Physics and Technology, School of Physics, Peking University, Beijing, 100871, China}

\author{Liangwen \surname{Chen}}
\affiliation{Institute of Modern Physics, Chinese Academy of Science, Lanzhou 730000, China; University of Chinese Academy of Sciences, Beijing 100049, China}

\author{Xueheng \surname{Zhang}}
\affiliation{Institute of Modern Physics, Chinese Academy of Science, Lanzhou 730000, China; University of Chinese Academy of Sciences, Beijing 100049, China}

\author{Zhiyu \surname{Sun}}
\affiliation{Institute of Modern Physics, Chinese Academy of Science, Lanzhou 730000, China; University of Chinese Academy of Sciences, Beijing 100049, China}

\author{Qiang \surname{Li}}
\email[]{qliphy0@pku.edu.cn}
\affiliation{State Key Laboratory of Nuclear Physics and Technology, School of Physics, Peking University, Beijing, 100871, China}

\begin{abstract}
Entanglement is a fundamental pillar of quantum mechanics. Probing quantum entanglement and testing Bell inequality with muons can be a significant leap forward, as muon is arguably the only massive elementary particle that can be manipulated and detected over a wide range of energies, e.g., from approximately 0.3 to $10^2$ GeV, corresponding to velocities from 0.94 to nearly the speed of light. In this work, we present a realistic proposal and a comprehensive study of quantum entanglement in a state composed of different-flavor fermions in muon-electron scattering. The polarization density matrix for the muon-electron system is derived using a kinematic approach within the relativistic quantum field theory framework. Entanglement in the resulting muon-electron qubit system and the violation of Bell inequalities can be observed with a high event rate. This paves the way for performing quantum tomography with muons.
\end{abstract}

\maketitle


\section{Introduction}

Quantum entanglement~\cite{Einstein:1935rr} is one of the most distinctive and counter-intuitive features of quantum mechanics. It describes a correlated quantum system that cannot be explained by classical theories, leading to the violation of Bell inequalities~\cite{Bell:1964kc}. Experimental searches for quantum entanglement and violations of Bell inequalities have been successfully performed using two-outcome measurements with correlated photon pairs~\cite{Freedman:1972zza,Clauser:1978ng,Aspect:1982fx}, thereby confirming the non-locality of quantum mechanics.

The theoretical framework for establishing quantum entanglement in quantum field theory (QFT) was introduced nearly two decades ago~\cite{Shi:2004yt}. Since then, various theoretical approaches have been developed to construct appropriate observables that quantify quantum entanglement within the context of QFT. For instance, quantum tomography methods have been applied to the production of top-quark and massive gauge-boson pair at the LHC, enabling the reconstruction of the density matrix from experimental data~\cite{Maina:2020rgd,Aguilar-Saavedra:2022mpg,Aguilar-Saavedra:2022wam,Aoude:2023hxv,Barr:2021zcp,Bi:2023uop,Fabbri:2023ncz,Marzola:2023oyv,Morales:2023gow,Grabarczyk:2024wnk,Afik:2020onf,Severi:2021cnj,Larkoski:2022lmv,Aguilar-Saavedra:2022uye,Afik:2022dgh,Afik:2022kwm,Han:2023fci,Dong:2023xiw}. Additionally, several proposals have been made to probe entanglement and test Bell inequality at future colliders~\cite{Ehataht:2023zzt,Ma:2023yvd,Gray:2021jij,Fabbrichesi:2023cev,Wu:2024ovc,Ruzi:2024cbt}.

Building on these theoretical approaches, the ATLAS and CMS Collaborations at the LHC have observed quantum entanglement between top and anti-top quarks~\cite{ATLAS:2023fsd,CMS:2024pts,CMS:2024zkc}. This achievement represents the most up-to-date measurement of quantum entanglement at the highest energy scale currently accessible by a particle collider.

Confirming the presence of quantum entanglement in high-energy particle physics not only enhances our fundamental understanding of quantum mechanics under extreme conditions but also paves the way for advancements in cutting-edge quantum technologies such as quantum computing, quantum simulation, and quantum teleportation. In addition to current achievements, probing quantum entanglement and testing Bell inequality with muons can be a significant leap forward, as muon is arguably the only massive elementary particle that can be manipulated and detected over a wide range of energies, e.g., from approximately 0.3 to $10^2$ GeV, corresponding to velocities from 0.94 to nearly the speed of light. This makes the muon probably the best candidate for testing the dependence of quantum entanglement on the velocity of the particles.

Attempts to probe quantum entanglement and test Bell inequality in highly relativistic particle collisions have mostly focused on decaying particles, such as massive gauge bosons and heavy fermions. The spin information of these particles is entirely carried by their decay products due to the parity-violating nature of the electroweak interaction, although challenging to extract due to the complexity of the production and decay processes. While for stable particles in a bipartite system, Refs.~\cite{Fedida:2022izl, Cheng:2024rxi} provide a simple yet effective \emph{kinematic approach} to handle the case where the outgoing particles can be directly detected at a collider.

\begin{figure}[t]
    \centering
    \includegraphics[width=0.9\linewidth]{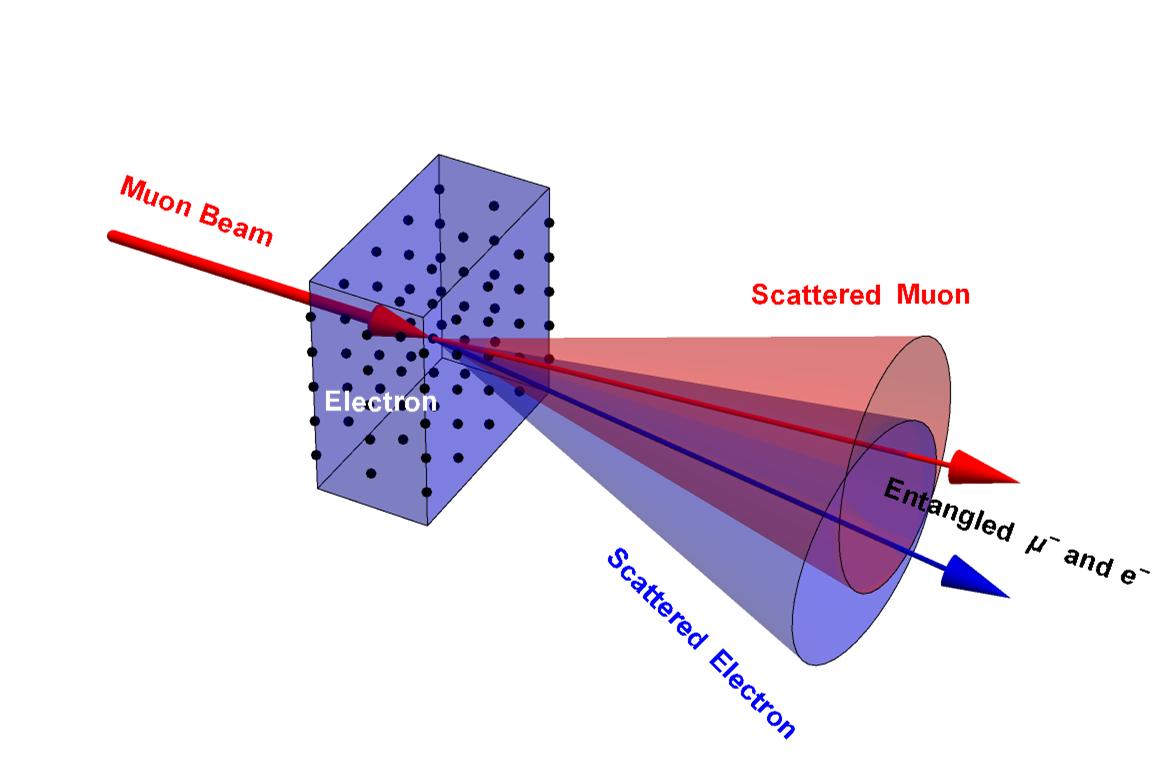}
    \caption{\justifying The schematic diagram showing a general muon on-target experiment.}
    \label{fig:bpt}
\end{figure}

In this letter, we investigate the presence of quantum entanglement and the violation of Bell inequalities in a muon-electron system through simulation of muon on-target experiments as illustrated in Fig.~\ref{fig:bpt}, partially referencing the PKMuon experiment proposals~\cite{Yu:2024spj, Gao:2024xvf}. In this setup, a muon beam with an energy of 0.3--$10^2$ GeV hits an electron approximately at rest in a target. The aforementioned kinematic approach is employed to derive the analytic final state density matrix. Throughout this work, we consider only the photon-exchange process at the tree level.

\section{Theoretical Framework}

As proposed in Refs.~\cite{Fedida:2022izl,Cheng:2024rxi}, the density matrix of the outgoing particles can be obtained as a function of kinematic quantities, such as the scattering angle relative to the incoming particle beam and the final particle's energy or momentum. Assuming the initial muon and electron states are unpolarized, the final state density matrix $\rho_f$ can be derived as
\begin{equation}
\rho_{s^\prime_3s^\prime_4s_3s_4} = \frac{1}{4}\sum_{s_1s_2}\frac{\mathcal{M}_{s^\prime_3s^\prime_4s_1s_2}\mathcal{M}^*_{s_3s_4s_1s_2}}{\sum_{s^{\prime\prime}_3s^{\prime\prime}_4}\left|\mathcal{M}_{s^{\prime\prime}_3s^{\prime\prime}_4s_1s_2}\right|^2},
\end{equation}
where $s_1$ and $s_2$ are the helicities of the muon and electron before scattering, and $s_3$ and $s_4$ are those after scattering. The polarized scattering amplitude $\mathcal{M}$ is calculated using the helicity spinors of each particle with momentum $p_i$ as
\begin{equation}
\begin{split}
\mathrm i\mc M_{s_3s_4s_1s_2} = \bar u(p_3, s_3)(-\mathrm ie\gamma^\mu)u(p_1, s_1) & \\
\frac{(-\mathrm ig_{\mu\nu})}{(p_1 - p_3)^2}\bar u(p_4, s_4)(-\mathrm ie\gamma^\nu)u(p_2, s_2)&.
\end{split}
\end{equation}

Entanglement can be quantified by several quantities~\cite{Fedida:2022izl}, known as a class of non-negative functions called entanglement monotone~\cite{Horodecki:2009zz,Chitambar:2018rnj}. One such quantity is \emph{concurrence}~\cite{Wootters:1997id,Gingrich:2002ota,Hill:1997pfa}. For a mixed state of two qubits, the concurrence is defined as
\begin{equation}
\mc{C}(\rho_f) = \max\{0, \lambda_1 - \lambda_2 - \lambda_3 - \lambda_4\},
\end{equation}
where $\lambda_i$ ($\lambda_i
\geq 
\lambda_j,\ \forall i < j$) are the square roots of the eigenvalues of matrix $\rho_f(\sigma_2 \otimes \sigma_2)\rho_f^*(\sigma_2 \otimes \sigma_2)$. If $\mc{C} > 0$, the two-qubit system is entangled. While $\mathcal{C}(\rho_f)$ is generally reference-frame-dependent~\cite{Gingrich:2002ota}, $\mathcal{C}(\rho_f)$ here inherits Lorentz-invariance from the scattering amplitude $\mathcal{M}$.

In addition, we can test the violation of the \emph{CHSH inequality}, $I_2 \leq 2$ \cite{Clauser:1969ny}, the Bell inequality for a two-qubit system, by evaluating the optimal (maximal) $I_2$ \cite{Horodecki:1995nsk} as
\begin{equation}
I_2 = 2\sqrt{\lambda_1 + \lambda_2},
\end{equation}
where $\lambda_1$ and $\lambda_2$ are the largest two eigenvalues of matrix $C^\mathrm{T}C$, and $C$ is the correlation matrix calculated by $C_{ij} = \text{Tr}\left(\rho_f\left(\sigma_i \otimes \sigma_j\right)\right)$. Any value of $I_2$ exceeding 2 indicates a strong evidence of quantum entanglement. The impact of statistical bias of the optimal CHSH value discussed in Ref.~\cite{Severi:2021cnj} is examined using large Monte Carlo samples in the following section.

\section{Simulation and Results}

\begin{figure*}[t]
\centering
\subfloat[center of mass frame]{\includegraphics[width=.45\linewidth]{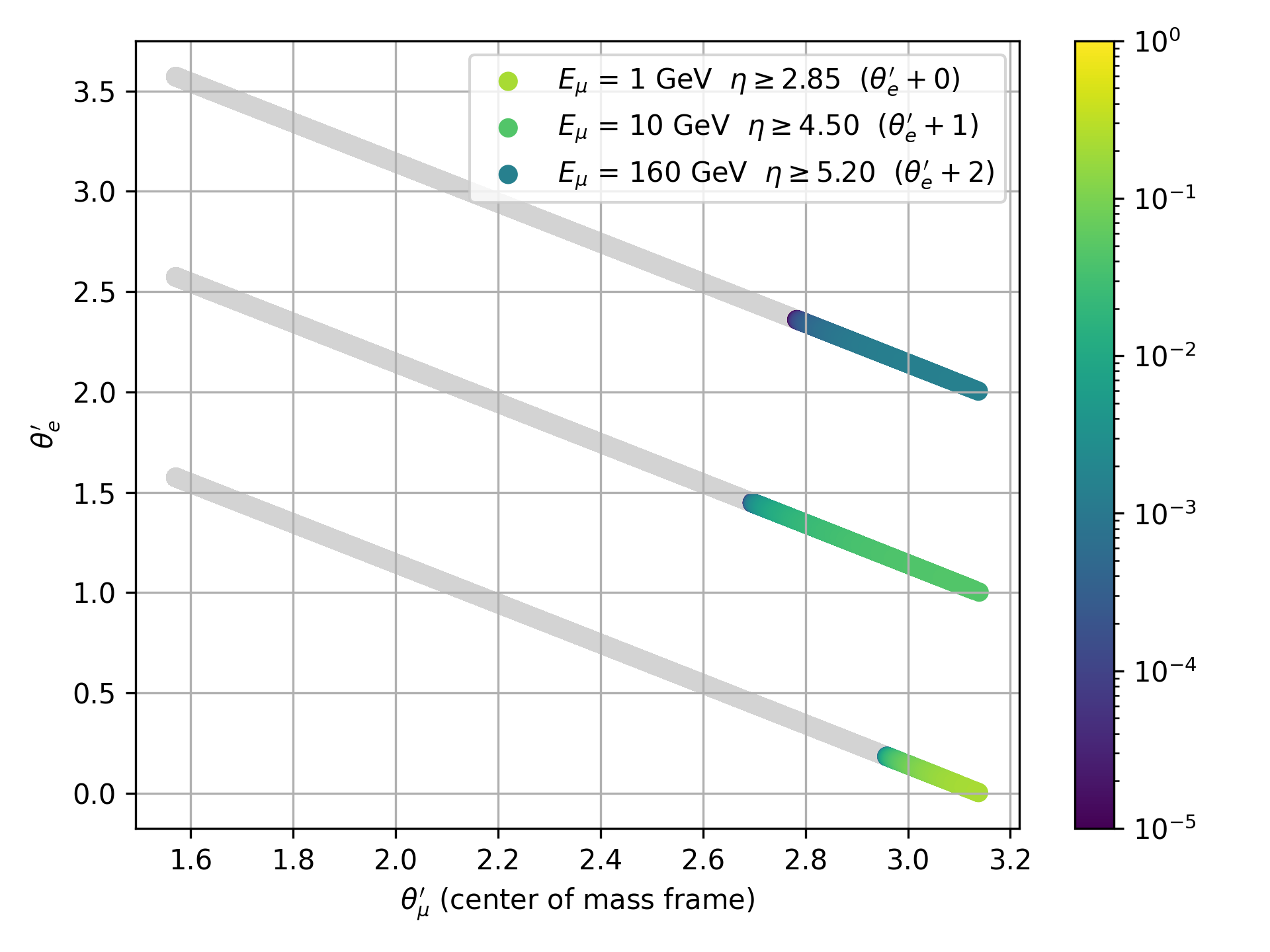}}
\subfloat[lab frame: $1\ \mathrm{GeV}$]{\includegraphics[width=.45\linewidth]{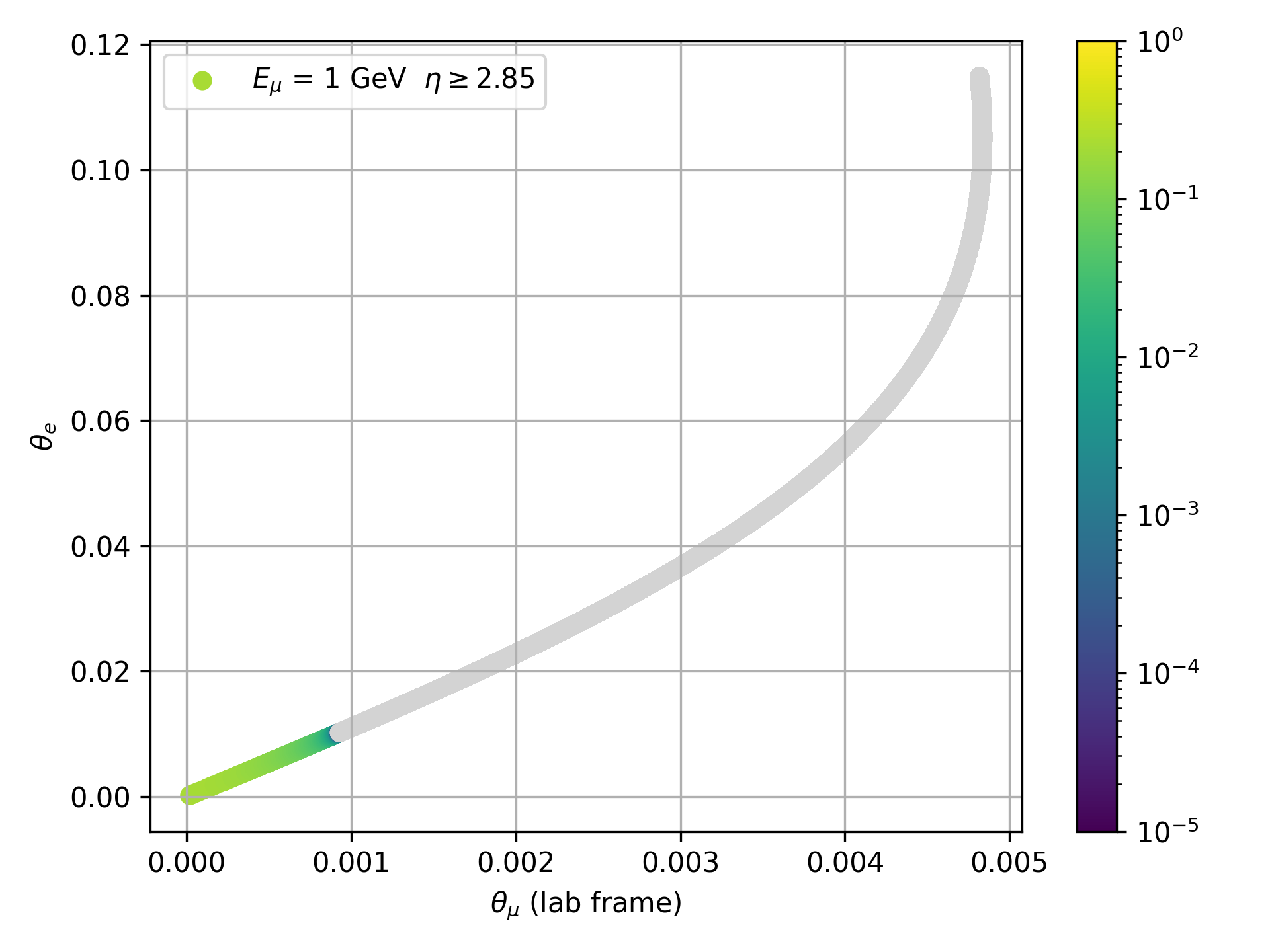}}

\subfloat[lab frame: $10\ \mathrm{GeV}$]{\includegraphics[width=.45\linewidth]{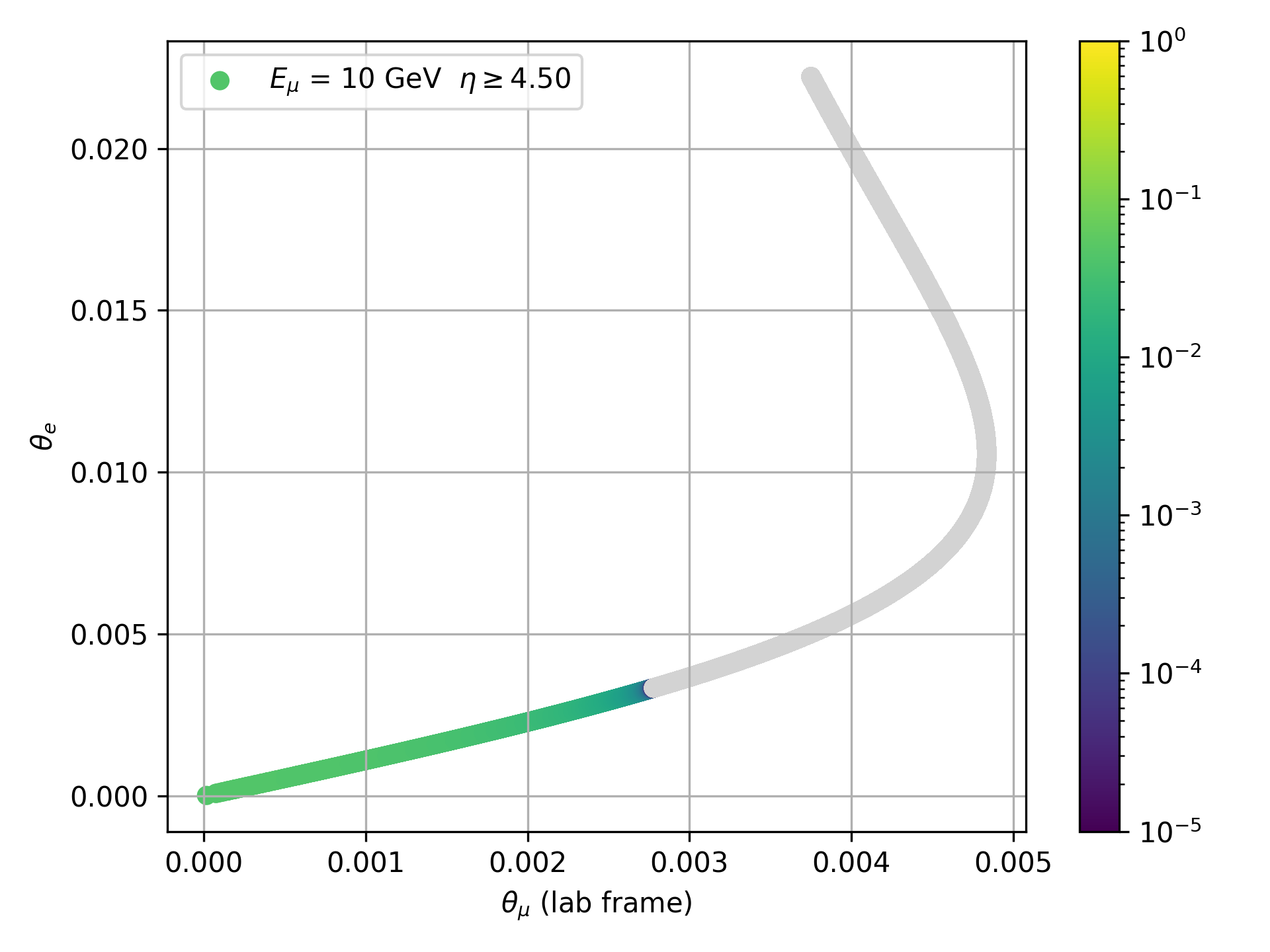}}
\subfloat[lab frame: $160\ \mathrm{GeV}$]{\includegraphics[width=.45\linewidth]{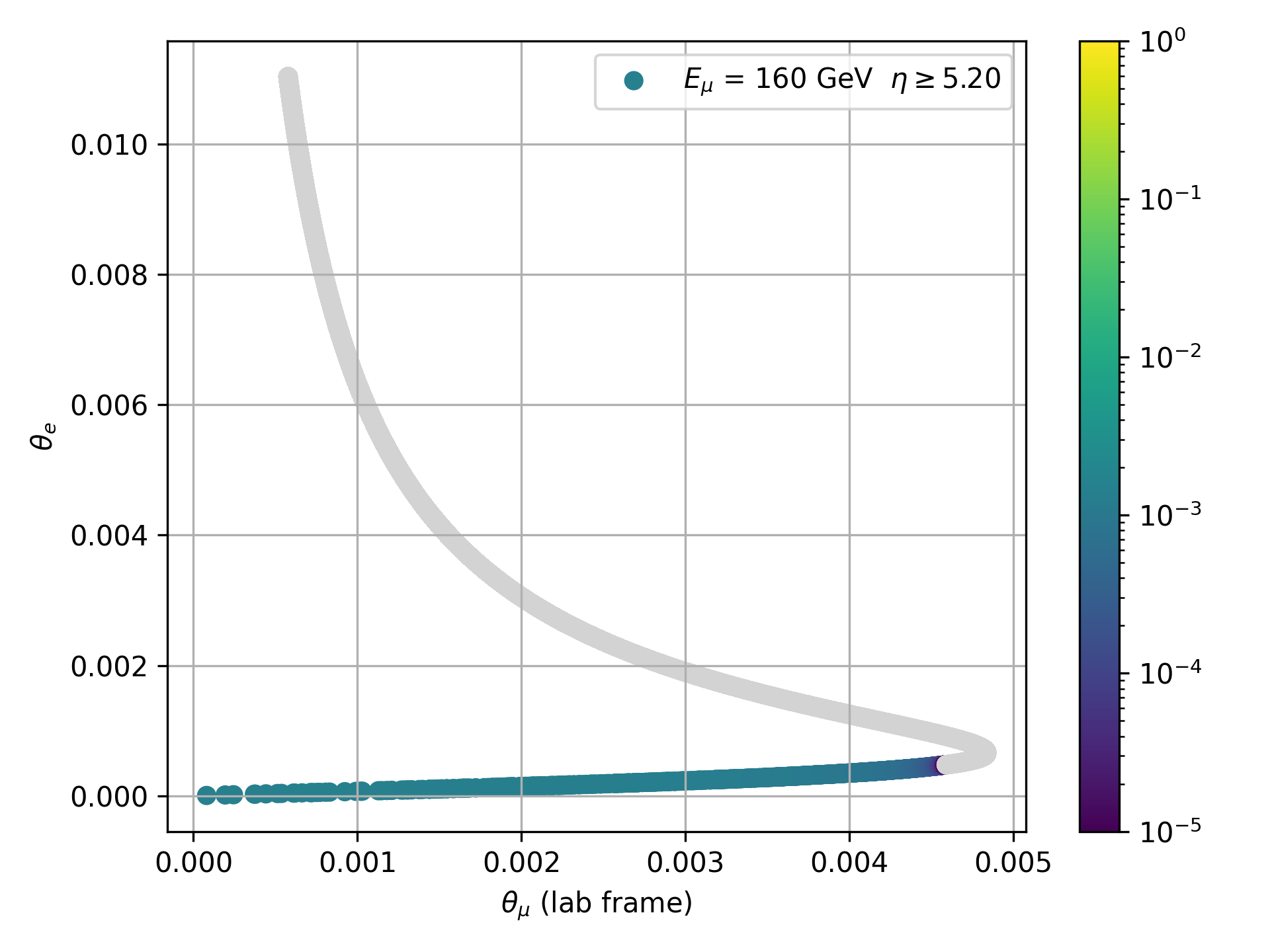}}
\caption{\justifying Final state scatter plots colored by concurrence $\mc{C}(\rho_f)$ for various incoming muon energies in (a) the center of mass frame and (b)--(d) the lab frame. Part of the unentangled phase space where $\theta^\prime_\mu \in [0, \pi/2)$ is omitted in subfigure (a).
The light gray regions depict $\mc C(\rho_f) = 0$.
Events are generated with the labeled minimum pseudorapidity requirements.}
\label{fig:qe}
\end{figure*}

The tree-level muon-electron scattering process is simulated by MadGraph5\_aMC@NLO 3.5.5~\cite{Alwall:2014hca} with non-zero lepton masses. Muon beams with energies of 1, 10, and 160 GeV are considered. Let $\theta_e$ ($\theta^\prime_e$) and $\theta_\mu$ ($\theta^\prime_\mu$) denote the polar angles of the final state electron and muon momenta in the lab (center of mass) frame. In the considered energy range, spin entanglement always occurs when $\theta^\prime_\mu$ approaches $\pi$, in agreement with Ref.~\cite{Fedida:2022izl}. Requirements on the final state lepton pseudorapidity ($\eta$, $\eta = -\ln\left[\tan\left(\theta/2\right)\right]$) are applied in the simulation to filter out the vast unentangled phase spaces with large $\theta_e$ or $\theta_\mu$ values.

The scatter plots in Fig.~\ref{fig:qe} depict concurrence $\mathcal{C}(\rho_f)$ distributions in terms of $\theta_{\mu}$ and $\theta_e$. The large margins with $\mathcal{C}(\rho_f) = 0$ shown in light gray exclude the possible dependence of the results on the applied pseudorapidity requirements. Stronger entanglement is anticipated at lower beam energies like 1 or 10 GeV, paving the way for probing entanglement with a cost-effective experimental setup. For the 1 GeV incoming muons, $\mc{C}(\rho_f)$ can exceed 0.2. In the 10 GeV case, both polar angles can approach 3 mrad simultaneously, making it feasible for detection and separation from the beam background. For the 160 GeV muon beam, weak entanglement is still present in the final state. However, the small polar angles due to the large Lorentz boost make it challenging for measurements.

\begin{table*}[t]
\caption{\justifying Summary of the kinematic variables for entangled phase space in the lab frame for various incident beam energies.}
\label{tab:summary}
\setlength{\tabcolsep}{.53em}
\begin{tabularx}{\linewidth}{rrrrrrrrrr}
\hline\hline
$E_\mathrm{beam}$/GeV &
$E_\mathrm{COM}$/GeV &
$\mc{C}(\rho_f)_{\max}$ &
$\theta_{\mu,\max}$/mrad &
$\theta_{e,\max}$/mrad &
$E_{\mu,\min}$/GeV &
$E_{e,\min}$/GeV &
$\sigma_{\mathrm E}$/\textmu b &
$\sigma_{\mathrm E,\,\theta \geq 0.5\,\mathrm{mrad}}$/\textmu b \\
\hline
1 & 0.111 & 0.22 & 0.9 & 10.2 & 0.92 & 0.08 &
0.56 & 
0.56 \\ 
10 & 0.146 & 0.044 & 2.8 & 3.3 & 5.2 & 4.5 &
0.39 & 
0.39 \\ 
160 & 0.418 & 0.0014 & 4.6 & 0.5 & 10 & 145 &
0.027 & 
0.022 \\ 
\hline\hline
\end{tabularx}
\end{table*}

\begin{figure*}[t]
\centering
\subfloat[center of mass frame]{\label{fig:CHSH-COM}\includegraphics[width=.45\linewidth]{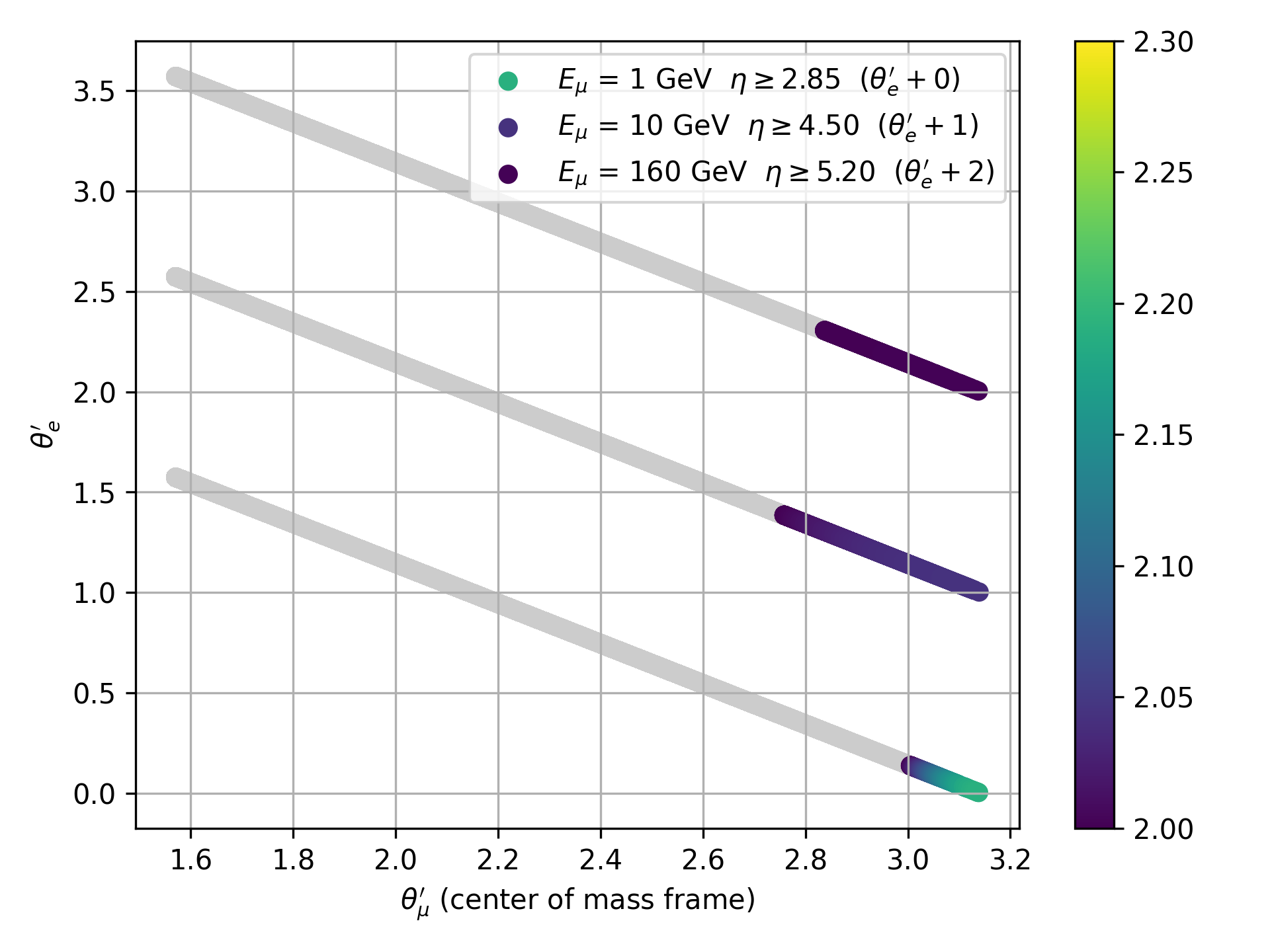}}
\subfloat[lab frame: $10\ \mathrm{GeV}$]{\label{fig:CHSH-10GeV}\includegraphics[width=.45\linewidth]{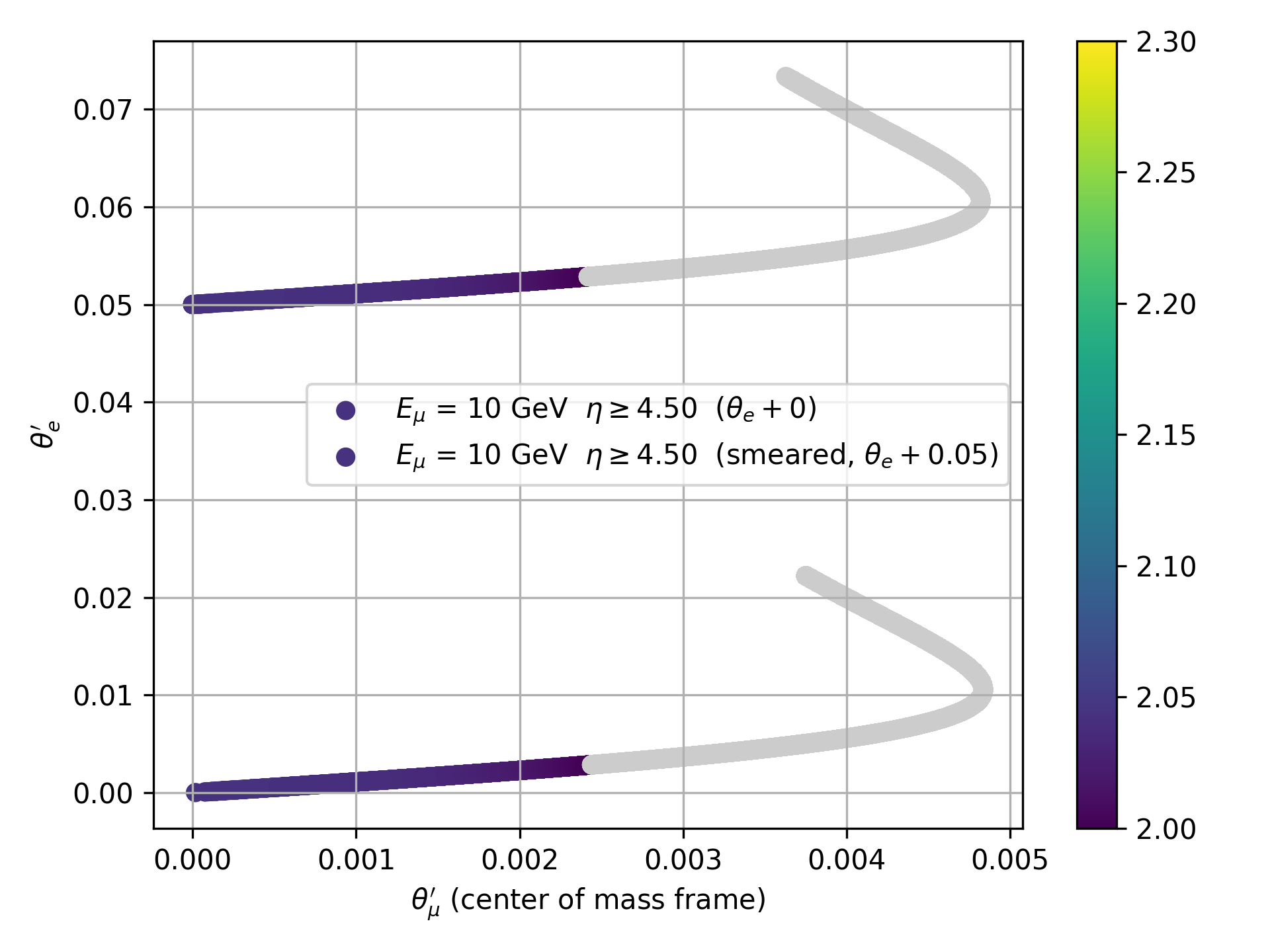}}
\caption{\justifying Final state scatter plots colored by the optimal CHSH value $I_2$ for various incoming muon energies in (a) the center of mass frame and (b) the lab frame. The light gray regions depict $I_2 \leq 2$.}
\label{fig:CHSH}
\end{figure*}

To analyze the expected yields of entangled events in future muon on-target experiments, we define the entangled cross section $\sigma_{\mathrm{E}}$ as the scattering cross section $\sigma$ multiplied by the ratio of events with $\mc{C}(\rho_f) > 0$. Similarly, $\sigma_{\mathrm E,\,\theta \geq 0.5\,\mathrm{mrad}}$ is defined for the subset of entangled events satisfying $\min\{\theta_\mu, \theta_e\} \geq 0.5\ \mathrm{mrad}$. Tab.~\ref{tab:summary} summarizes the results obtained for the three different beam energies. While entanglement can generally be measured across a wide energy range, the 10 GeV configuration offers balanced indicators and imposes minimal requirements on the detectors. To illustrate the high production rate of entangled events in an on-target experiment, assuming a one-day run with a 10 GeV muon beam of flux $10^5/\mathrm{s}$ on aluminum targets stacked up to $10\ \mathrm{cm}$ thick, the expected number of events with entangled final states is $2.6 \times 10^{4}$.  

In addition to the raw simulation results, we estimate the impact of measurement uncertainties in a real experiment by randomly smearing $\theta_\mu$, $\theta_e$, $E_\mu$, and $E_e$ with Gaussian errors of 0.3 mrad for each angle and 0.5 GeV for each energy. Since these four observables are functions of $\theta^\prime_\mu$, we fit $\theta^\prime_\mu$ to minimize the summed square relevant errors $\chi^2$ of the four observables and update their values according to the best fit $\theta^\prime_\mu$. In $10^6$ Monte Carlo events generated for the 10 GeV incoming muons, using the smeared-and-fit observables, we find that the minimized $\chi^2$ follows a $\chi^2$ distribution with degree of freedom of $2.99 \approx 4 - 1$. By removing events with $\mc{C}(\rho_f) \leq 10^{-5}$, we achieve a true positive rate\footnote{Given a threshold $\mc{C}(\rho_f) = \mc{C}_0$, the true positive rate is defined as the ratio of events where $\mc{C}(\rho_f) > \mc{C}_0$ after smearing to all events where $\mc{C}(\rho_f) > 0$ before smearing, and the false positive rate as the ratio of where $\mc{C}(\rho_f) > \mc{C}_0$ after to all where $\mc{C}(\rho_f) = 0$ before.} of 0.94 with a false positive rate upper limit of $10^{-3}$. This reveals the robustness of $\mc{C}(\rho_f)$ as the criterion of entanglement under the considered detector resolution.

Fig.~\ref{fig:CHSH-COM} shows the phase spaces where the CHSH inequality is violated. They are proper subsets of the entangled phase spaces shown in Fig.~\ref{fig:qe}, containing 0.52, 0.71, and 0.76 times the number of events in Fig.~\ref{fig:qe} for the 1, 10, and 160 GeV muon beams, respectively. That means one can also observe the violation of the CHSH inequality with the same experiment measuring quantum entanglement. As with the $\mc{C}(\rho_f)$ results, we also use the smeared $10^6$ Monte Carlo events to calculate $I_2$ for the 10 GeV muon beam and visualize the results in Fig.~\ref{fig:CHSH-10GeV}. The plot indicates that the bias is negligible with large statistics.

\section{Conclusion and Outlook}

Among the elementary particles in the Standard Model, the muon stands out as virtually the only candidate for testing the dependence of quantum entanglement on a wide range of particle energies and velocities. In this work, a novel kinematic approach is applied to the non-decaying, directly detected final state particles in the tree-level muon-electron scattering process. Through Monte Carlo simulation, quantum entanglement is found to be observable in muon on-target experiments with incident beam energies of 1, 10, and 160 GeV. The 10 GeV configuration imposes minimal requirements on the detectors and can achieve a high event rate ($\sim 10^4/\mathrm{d}$) meanwhile with a low misidentification rate ($\sim 10^{-3}$) due to measurement uncertainties. The violation of the CHSH inequality can be observed in a large subset (52--76\%) of the entangled events.

High-energy muon beams, such as the 160 GeV beam available at CERN’s MUonE experiment~\cite{CarloniCalame:2015obs}, and beams in the 1–10 GeV range, soon to be accessible at facilities like China’s HIAF~\cite{Yang:2013yeb}, provide an ideal platform for these studies. This work, together with the increasing availability of muon beams globally, lays a solid foundation for performing quantum state tomography and exploring advanced topics in quantum physics using muons.

\section*{Acknowledgments}

This work is supported in part by the National Natural Science Foundation of China under Grants No. 12150005, No. 12325504, and No. 12075004.


\bibliographystyle{ieeetr}
\bibliography{ref}

\begin{thebibliography}{10}

\bibitem{Einstein:1935rr}
A.~Einstein, B.~Podolsky, and N.~Rosen, ``{Can quantum mechanical description of physical reality be considered complete?},'' {\em Phys. Rev.}, vol.~47, pp.~777--780, 1935.

\bibitem{Bell:1964kc}
J.~S. Bell, ``{On the Einstein-Podolsky-Rosen paradox},'' {\em Physics Physique Fizika}, vol.~1, pp.~195--200, 1964.

\bibitem{Freedman:1972zza}
S.~J. Freedman and J.~F. Clauser, ``{Experimental Test of Local Hidden-Variable Theories},'' {\em Phys. Rev. Lett.}, vol.~28, pp.~938--941, 1972.

\bibitem{Clauser:1978ng}
J.~F. Clauser and A.~Shimony, ``{Bell's theorem: Experimental tests and implications},'' {\em Rept. Prog. Phys.}, vol.~41, pp.~1881--1927, 1978.

\bibitem{Aspect:1982fx}
A.~Aspect, J.~Dalibard, and G.~Roger, ``{Experimental test of Bell's inequalities using time varying analyzers},'' {\em Phys. Rev. Lett.}, vol.~49, pp.~1804--1807, 1982.

\bibitem{Shi:2004yt}
Y.~Shi, ``{Entanglement in relativistic quantum field theory},'' {\em Phys. Rev. D}, vol.~70, p.~105001, 2004.

\bibitem{Maina:2020rgd}
E.~Maina, ``{Vector boson polarizations in the decay of the Standard Model Higgs},'' {\em Phys. Lett. B}, vol.~818, p.~136360, 2021.

\bibitem{Aguilar-Saavedra:2022mpg}
J.~A. Aguilar-Saavedra, ``{Laboratory-frame tests of quantum entanglement in H\textrightarrow{}WW},'' {\em Phys. Rev. D}, vol.~107, no.~7, p.~076016, 2023.

\bibitem{Aguilar-Saavedra:2022wam}
J.~A. Aguilar-Saavedra, A.~Bernal, J.~A. Casas, and J.~M. Moreno, ``{Testing entanglement and Bell inequalities in H\textrightarrow{}ZZ},'' {\em Phys. Rev. D}, vol.~107, no.~1, p.~016012, 2023.

\bibitem{Aoude:2023hxv}
R.~Aoude, E.~Madge, F.~Maltoni, and L.~Mantani, ``{Probing new physics through entanglement in diboson production},'' {\em JHEP}, vol.~12, p.~017, 2023.

\bibitem{Barr:2021zcp}
A.~J. Barr, ``{Testing Bell inequalities in Higgs boson decays},'' {\em Phys. Lett. B}, vol.~825, p.~136866, 2022.

\bibitem{Bi:2023uop}
Q.~Bi, Q.-H. Cao, K.~Cheng, and H.~Zhang, ``{New observables for testing Bell inequalities in $W$ boson pair production},'' 7 2023.

\bibitem{Fabbri:2023ncz}
F.~Fabbri, J.~Howarth, and T.~Maurin, ``{Isolating semi-leptonic $H\rightarrow WW^{*}$decays for Bell inequality tests},'' {\em Eur. Phys. J. C}, vol.~84, no.~1, p.~20, 2024.

\bibitem{Marzola:2023oyv}
L.~Marzola, ``{Testing Bell inequalities and entanglement with di-boson final states},'' in {\em {57th Rencontres de Moriond on Electroweak Interactions and Unified Theories}}, 5 2023.

\bibitem{Morales:2023gow}
R.~A. Morales, ``{Exploring Bell inequalities and quantum entanglement in vector boson scattering},'' {\em Eur. Phys. J. Plus}, vol.~138, no.~12, p.~1157, 2023.

\bibitem{Grabarczyk:2024wnk}
R.~Grabarczyk, ``{An improved Bell-CHSH observable for gauge boson pairs},'' 10 2024.

\bibitem{Afik:2020onf}
Y.~Afik and J.~R. M.~n. de~Nova, ``{Entanglement and quantum tomography with top quarks at the LHC},'' {\em Eur. Phys. J. Plus}, vol.~136, no.~9, p.~907, 2021.

\bibitem{Severi:2021cnj}
C.~Severi, C.~D.~E. Boschi, F.~Maltoni, and M.~Sioli, ``{Quantum tops at the LHC: from entanglement to Bell inequalities},'' {\em Eur. Phys. J. C}, vol.~82, no.~4, p.~285, 2022.

\bibitem{Larkoski:2022lmv}
A.~J. Larkoski, ``{General analysis for observing quantum interference at colliders},'' {\em Phys. Rev. D}, vol.~105, no.~9, p.~096012, 2022.

\bibitem{Aguilar-Saavedra:2022uye}
J.~A. Aguilar-Saavedra and J.~A. Casas, ``{Improved tests of entanglement and Bell inequalities with LHC tops},'' {\em Eur. Phys. J. C}, vol.~82, no.~8, p.~666, 2022.

\bibitem{Afik:2022dgh}
Y.~Afik and J.~R. M.~n. de~Nova, ``{Quantum Discord and Steering in Top Quarks at the LHC},'' {\em Phys. Rev. Lett.}, vol.~130, no.~22, p.~221801, 2023.

\bibitem{Afik:2022kwm}
Y.~Afik and J.~R. M.~n. de~Nova, ``{Quantum information with top quarks in QCD},'' {\em Quantum}, vol.~6, p.~820, 2022.

\bibitem{Han:2023fci}
T.~Han, M.~Low, and T.~A. Wu, ``{Quantum Entanglement and Bell Inequality Violation in Semi-Leptonic Top Decays},'' 10 2023.

\bibitem{Dong:2023xiw}
Z.~Dong, D.~Gon\c{c}alves, K.~Kong, and A.~Navarro, ``{Entanglement and Bell inequalities with boosted tt\textasciimacron{}},'' {\em Phys. Rev. D}, vol.~109, no.~11, p.~115023, 2024.

\bibitem{Ehataht:2023zzt}
K.~Ehat\"aht, M.~Fabbrichesi, L.~Marzola, and C.~Veelken, ``{Probing entanglement and testing Bell inequality violation with $\textrm{e}^{+}\textrm{e}^{-} \rightarrow \tau^{+}\tau^{-}$ at Belle II},'' 11 2023.

\bibitem{Ma:2023yvd}
K.~Ma and T.~Li, ``{Testing Bell inequality through $h\to\tau\tau$ at CEPC},'' 9 2023.

\bibitem{Gray:2021jij}
H.~M. Gray, ``{Future colliders for the high-energy frontier},'' {\em Rev. Phys.}, vol.~6, p.~100053, 2021.

\bibitem{Fabbrichesi:2023cev}
M.~Fabbrichesi, R.~Floreanini, E.~Gabrielli, and L.~Marzola, ``{Bell inequalities and quantum entanglement in weak gauge boson production at the LHC and future colliders},'' {\em Eur. Phys. J. C}, vol.~83, no.~9, p.~823, 2023.

\bibitem{Wu:2024ovc}
Y.~Wu, R.~Jiang, A.~Ruzi, Y.~Ban, and Q.~Li, ``{Testing Bell inequalities and probing quantum entanglement at CEPC},'' 10 2024.

\bibitem{Ruzi:2024cbt}
A.~Ruzi, Y.~Wu, R.~Ding, S.~Qian, A.~M. Levin, and Q.~Li, ``{Testing Bell inequalities and probing quantum entanglement at a muon collider},'' {\em JHEP}, vol.~10, p.~211, 2024.

\bibitem{ATLAS:2023fsd}
G.~Aad {\em et~al.}, ``{Observation of quantum entanglement with top quarks at the ATLAS detector},'' {\em Nature}, vol.~633, no.~8030, pp.~542--547, 2024.

\bibitem{CMS:2024pts}
A.~Hayrapetyan {\em et~al.}, ``{Observation of quantum entanglement in top quark pair production in proton\textendash{}proton collisions at $\sqrt{s} = 13$ TeV},'' {\em Rept. Prog. Phys.}, vol.~87, no.~11, p.~117801, 2024.

\bibitem{CMS:2024zkc}
A.~Hayrapetyan {\em et~al.}, ``{Measurements of polarization and spin correlation and observation of entanglement in top quark pairs using lepton+jets events from proton-proton collisions at $\sqrt{s}$ = 13 TeV},'' 9 2024.

\bibitem{Fedida:2022izl}
S.~Fedida and A.~Serafini, ``{Tree-level entanglement in quantum electrodynamics},'' {\em Phys. Rev. D}, vol.~107, no.~11, p.~116007, 2023.

\bibitem{Cheng:2024rxi}
K.~Cheng, T.~Han, and M.~Low, ``{Quantum Tomography at Colliders: With or Without Decays},'' 10 2024.

\bibitem{Yu:2024spj}
X.~Yu {\em et~al.}, ``{Proposed Peking University muon experiment for muon tomography and dark matter search},'' {\em Phys. Rev. D}, vol.~110, no.~1, p.~016017, 2024.

\bibitem{Gao:2024xvf}
L.~Gao, Z.~Wang, C.-e. Liu, J.~Li, A.~Ruzi, Q.~Li, C.~Zhou, and Q.~Li, ``{Probing charged lepton flavor violation in an economical muon on-target experiment},'' 10 2024.

\bibitem{Horodecki:2009zz}
R.~Horodecki, P.~Horodecki, M.~Horodecki, and K.~Horodecki, ``{Quantum entanglement},'' {\em Rev. Mod. Phys.}, vol.~81, pp.~865--942, 2009.

\bibitem{Chitambar:2018rnj}
E.~Chitambar and G.~Gour, ``{Quantum resource theories},'' {\em Rev. Mod. Phys.}, vol.~91, no.~2, p.~025001, 2019.

\bibitem{Wootters:1997id}
W.~K. Wootters, ``{Entanglement of formation of an arbitrary state of two qubits},'' {\em Phys. Rev. Lett.}, vol.~80, pp.~2245--2248, 1998.

\bibitem{Gingrich:2002ota}
R.~M. Gingrich and C.~Adami, ``{Quantum Entanglement of Moving Bodies},'' {\em Phys. Rev. Lett.}, vol.~89, p.~270402, 2002.

\bibitem{Hill:1997pfa}
S.~Hill and W.~K. Wootters, ``{Entanglement of a pair of quantum bits},'' {\em Phys. Rev. Lett.}, vol.~78, pp.~5022--5025, 1997.

\bibitem{Clauser:1969ny}
J.~F. Clauser, M.~A. Horne, A.~Shimony, and R.~A. Holt, ``{Proposed experiment to test local hidden variable theories},'' {\em Phys. Rev. Lett.}, vol.~23, pp.~880--884, 1969.

\bibitem{Horodecki:1995nsk}
R.~Horodecki, P.~Horodecki, and M.~Horodecki, ``{Violating Bell inequality by mixed spin-1/2 states: necessary and sufficient condition},'' {\em Phys. Lett. A}, vol.~200, no.~5, pp.~340--344, 1995.

\bibitem{Alwall:2014hca}
J.~Alwall, R.~Frederix, S.~Frixione, V.~Hirschi, F.~Maltoni, O.~Mattelaer, H.~S. Shao, T.~Stelzer, P.~Torrielli, and M.~Zaro, ``{The automated computation of tree-level and next-to-leading order differential cross sections, and their matching to parton shower simulations},'' {\em JHEP}, vol.~07, p.~079, 2014.

\bibitem{CarloniCalame:2015obs}
C.~M. Carloni~Calame, M.~Passera, L.~Trentadue, and G.~Venanzoni, ``{A new approach to evaluate the leading hadronic corrections to the muon $g$-2},'' {\em Phys. Lett. B}, vol.~746, pp.~325--329, 2015.

\bibitem{Yang:2013yeb}
J.~C. Yang {\em et~al.}, ``{High Intensity heavy ion Accelerator Facility (HIAF) in China},'' {\em Nucl. Instrum. Meth. B}, vol.~317, pp.~263--265, 2013.

\end{thebibliography}

\end{document}